# Intermolecular distance and density scaling of dynamics in molecular liquids

D. Fragiadakis and C. M. Roland

*Naval Research Laboratory, Chemistry Division, Washington, DC 20375-5342, USA*

(April 2, 2019)

**ABSTRACT**

A broad variety of liquids conform to density scaling: relaxation times expressed as a function of the ratio of temperature to density, the latter raised to a material constant $\gamma$. For atomic liquids interacting only through simple pair potentials, the exponent $\gamma$ is very nearly equal to $n/3$, where $n$ is the steepness of the intermolecular potential, while for molecular liquids having rigid bonds and built using the same interatomic potential, $\gamma > n/3$. We find that for this class of molecular liquids $\gamma = n/\delta$, where the parameter $\delta$ relates the intermolecular distance to the density along an isomorph (line of approximately constant dynamics and structure). $\delta$ depends only on the molecular structure and not the interatomic potential.

**INTRODUCTION**

The dynamics of viscous liquids is very sensitive to temperature and pressure. Close to the glass transition, small changes in temperature or pressure can change the relaxation time, viscosity and diffusion coefficient by many orders of magnitude. Furthermore, the temperature and pressure dependencies of the dynamics differ greatly among materials. We are still far from being able to predict liquid dynamics based on molecular structure—a fundamental understanding of this has long been a goal of condensed matter physics[1,2,3,4,5,6].

An important development in the understanding of the dynamics of supercooled liquids was the discovery of density scaling, the fact that the relaxation times $\tau$ and other dynamic quantities can be expressed as a function of the ratio of temperature and density, the latter raised to a material constant $\gamma$[7,8,9,10,11].

$$\tau = f(\rho^\gamma/T) \qquad (1)$$

This property has been verified for more than 100 liquids and polymers[12,13], with the latter generally having smaller γ. The only materials deviating from eq. 1 are ones that undergo changes in structure (such as degree of hydrogen bonding) with temperature and pressure. A related property is isochronal superpositioning: the shape of the relaxation spectrum, although it varies with state point, depends only on the relaxation time[14,15].

For a system of particles interacting through an inverse power law (IPL) pair potential $u(r) \propto r^{-n}$, both density scaling and isochronal superposition are exact, with $\gamma = n/3$, when quantities are expressed in reduced units[16,17].

$$l_0 = \rho^{1/3}, \; t_0 = \rho^{-1/3}\sqrt{m/k_B T}, \; \epsilon_0 = k_B T \tag{2}$$

The difference between scaling using reduced and unreduced units is negligible in the supercooled regime; however, at higher temperature the difference can be substantial[18]. Using computer simulations, a broad class of liquids has been discovered showing density scaling and isochronal superposition to a very good approximation. These systems have so-called isomorphs, lines of constant structure and dynamics, in reduced units, in their $(\rho, T)$ phase diagram, and show strong correlation between the fluctuations of the potential energy $\Delta U$ and virial $\Delta W$, with the proportionality constant being the density scaling exponent $\gamma$[19]. For simulated liquids with isomorphs, $\gamma$ in general varies with state point although it is to good approximation a function of density[20]. However, for reasons not completely understood, at least some real liquids are well described using a constant $\gamma$, even for large density changes[21,22,23]. Systems that have isomorphs have been called Roskilde-simple, or R-simple systems[24], and include atomic liquids and mixtures, but also simple molecular and polymeric liquids. Among real materials, non-associating liquids and polymers as well as ionic liquids are expected to be R-simple, while those with strong intermolecular hydrogen bonds are not [25]. However, quantifying $\Delta U$ and $\Delta W$ for real materials is challenging [26].

Based on the relation $\gamma = n/3$ for the inverse power law potential, it is natural to expect that for liquids that exhibit density scaling, the exponent $\gamma$ is in some way related to the steepness of the intermolecular potential. This relationship is clear for simulated atomic systems interacting through a pair potential $u(r)$; for these systems $\gamma$ can in fact be accurately estimated from the derivatives of $u(r)$ near the most likely interatomic distance[27,28]. However it is not yet known to what extent this relationship can be extended to molecular and polymeric liquids, where the intermolecular potential depends not only on distance but on the relative orientation of two molecules or chains and on their internal degrees of freedom. The key question is what is the relevant potential? Does there exist for these systems an "effective" intermolecular or intersegment pair potential $u_{eff}(r_{eff})$, related to $\gamma$ in the same way as $u(r)$ is in atomic systems, and if so, what are $u_{eff}$ and $r_{eff}$? Herein we attempt a first step toward this generalization, on simple molecules and polymer chains defined by pair potentials and rigid bonds, which are known to be R-simple.

**RESULTS AND DISCUSSION**

Simulations are carried out using the RUMD simulation software[29], all in the NVT ensemble with a Nose-Hoover thermostat[30]. We study three different interatomic potentials: the standard 12-6 Lennard-Jones potential[31] (LJ), and two inverse power laws with n=12 and n=18 (IPL12 and IPL18, respectively):

$$u_{LJ}(r) = 4\varepsilon\left[\left(\frac{r}{\sigma}\right)^{-12} - \left(\frac{r}{\sigma}\right)^{-6}\right] \tag{3}$$

**Table I.** Summary of the isomorph simulated for each liquid: minimum and maximum density and temperature, average correlation coefficient and scaling exponent (eq. 6), exponent $\delta$ relating intermolecular distance to density. Its product with the average (effective?) scaling exponent, in the last column is approximately equal to the steepness of the *interatomic* potential, which is the main result of this paper.

|  | Potential | $(\rho_0, T_0)$ | $(\rho_1, T_1)$ | $\bar{R}$ | $\bar{\gamma}$ | $\delta$ | $n = \bar{\gamma}\delta$ |
|---|---|---|---|---|---|---|---|
| Atomic | LJ | (1.00, 1.00) | (1.30, 3.58) | 0.99 | 4.96 | 3.00 | 14.9 |
|  | IPL12 | (1.00, 1.00) | (1.30, 2.86) | 1.00 | 4.00 | 3.00 | 12.0 |
|  | IPL18 | (1.00, 1.00) | (1.30, 4.83) | 1.00 | 6.00 | 3.00 | 18.0 |
| Asymmetric Dumbbell | LJ | (1.64, 0.29) | (2.14, 1.29) | 0.96 | 5.66 | 2.68 | 15.2 |
|  | IPL12 | (1.64, 0.12) | (2.14, 0.41) | 0.93 | 4.66 | 2.63 | 12.2 |
|  | IPL18 | (1.64, 0.17) | (2.14, 1.05) | 0.96 | 6.97 | 2.61 | 18.2 |
| Lewis-Wahnström OTP | LJ | (1.00, 0.78) | (1.30, 4.09) | 0.88 | 6.27 | 2.36 | 14.8 |
|  | IPL12 | (1.00, 0.60) | (1.30, 2.39) | 0.86 | 5.22 | 2.36 | 12.3 |
|  | IPL18 | (1.00, 0.70) | (1.30, 5.33) | 0.90 | 7.69 | 2.36 | 18.1 |
| Freely Jointed Chain N=10 | LJ | (1.00, 0.71) | (1.30, 3.90) | 0.83 | 6.49 | 2.28 | 14.8 |
|  | IPL12 | (1.00, 2.24) | (1.30, 9.14) | 0.82 | 5.35 | 2.22 | 11.9 |
|  | IPL18 | (1.00, 2.77) | (1.30, 22.6) | 0.85 | 7.98 | 2.26 | 18.0 |

$$u_{IPL12}(r) = \varepsilon \left(\frac{r}{\sigma}\right)^{-12} \tag{4}$$

$$u_{IPL18}(r) = \varepsilon \left(\frac{r}{\sigma}\right)^{-18} \tag{5}$$

For each potential we simulate the single-component atomic liquid, as well as three molecules: (a) an asymmetric dumbbell shaped molecule (ADB) [32,33,34]; (b) the Lewis-Wahnström *o*-terphenyl (LW-OTP) model, a rigid trimer with a bond angle constrained to 75 degrees[35,36]; and (c) a freely jointed chain of 10 atoms linked by rigid bonds[37]. All systems have been found in previous studies [33,37] to have isomorphs to good approximation.

Each system was simulated at a number of state points along an isomorph following the methodology of ref. [38]. Beginning at a state point $(\rho_0, T_0)$, the scaling exponent $\gamma$ and correlation coefficient $R$ were determined from the fluctuations of the potential energy $U$ and virial $W$

$$\gamma = \frac{\langle \Delta U \Delta W \rangle}{\langle (\Delta U)^2 \rangle} \tag{6}$$

$$R = \frac{\langle \Delta W \Delta U \rangle}{\sqrt{\langle (\Delta W)^2 \rangle \langle (\Delta U)^2 \rangle}} \tag{7}$$

To step to the next state point along the isomorph a small density change $\delta\rho$ was made along with the temperature change $\delta T$ required to remain on the isomorph defined by $\Delta \ln T = \gamma \Delta \ln \rho$. The process was then repeated to trace out an isomorph from $(\rho_0, T_0)$ to $(\rho_1, T_1)$, with a ~30% density change, of the same order as some of the largest density changes investigated experimentally (for example cumene[22], polyurea[21], and nitrogen [23]). The values of $\gamma$ and $R$ as a function of density along these isomorphs are shown in Fig. 1, with the initial state points and average values of $\gamma$ and $R$ given in Table I.

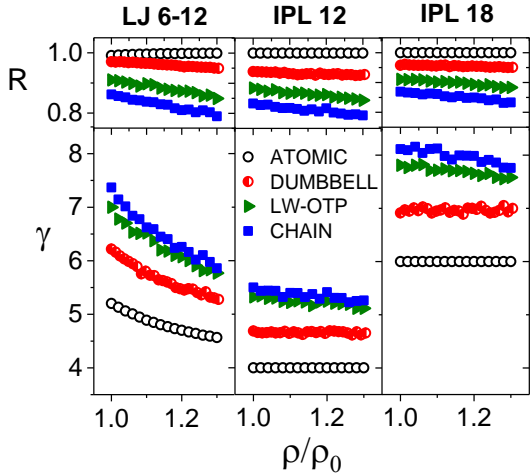
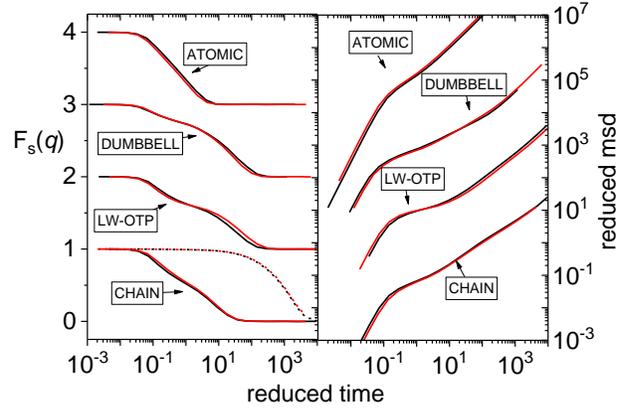

**Figure 1.** Scaling exponent and correlation coefficient determined from U-W correlations along the isomorphs of Table I, for atomic systems and the dumbbell, LW-OTP and chain molecules consisting of Lennard-Jones, IPL-12 and IPL-18 atoms.

**Figure 2.** Self-intermediate scattering function at a $q$ corresponding the maximum of the static structure factor for each state point (solid lines), and mean square displacement, at the two extreme state points of the isomorph of Table I for each of the four liquids based on LJ potential. Black: $(\rho_0, T_0)$, Red: $(\rho_1, T_1)$ For the chain molecule, the autocorrelation function of the end-to-end vector of the chains is also shown in the left panel (dashed line).

Dependence of $\gamma$ on molecular shape

The two atomic IPL systems with n=12 and n=18 have perfect pressure-energy correlations, i.e., R=1, and constant $\gamma = n/3$. The LJ system still has almost perfect UW correlations (R>0.99, decreasing very slightly with decreasing density), but the effective slope of the potential is steeper resulting in larger $\gamma$ which depends significantly on density.

For all three interatomic potentials $\gamma$ increases significantly going from atomic to molecular liquids, and for all three potentials the order from smallest to largest $\gamma$ is atomic, ADB, LW-OTP and finally chain molecules. Note that $\gamma$ is no longer constant for the IPL based molecules although its density dependence is quite weak. The molecular systems are also more weakly correlating (smaller R) than the corresponding atomic ones and R decreases in the same order that $\gamma$ increases. R also decreases with increasing density, the opposite behavior to that of the atomic LJ liquid.

As has been found previously [32–37], the dynamics along the isomorph, in reduced units, are essentially invariant (Fig. 2): both the mean square displacement and self-intermediate scattering function practically overlap at the highest and lowest densities. This means the difference between isomorphs and isochrones is very small, and in fact the following results do not qualitatively change if we use isochrones (by requiring constant relaxation time or constant mean square displacement, in reduced units) instead of tracing isomorphs.

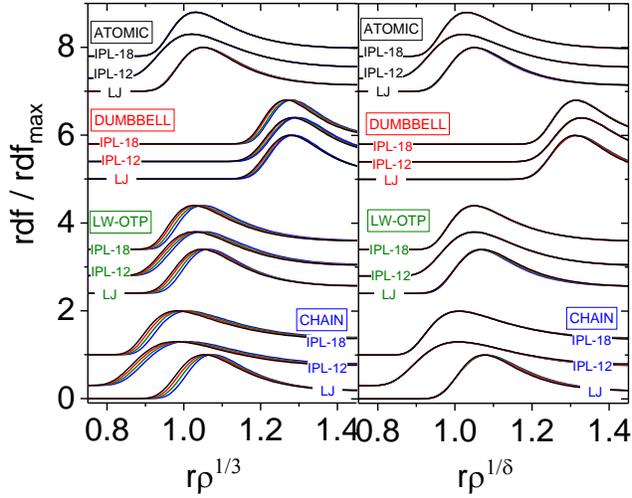
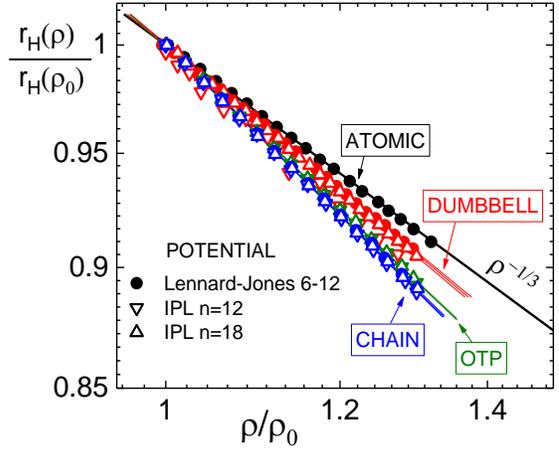

**Figure 3.** Intermolecular part of the radial distribution function for the simulated liquids, in the region around the first peak, plotted against reduced length, i.e., r scaled by $\rho^{-1/3}$ (left) and r scaled by $\rho^{-1/\delta}$ (right) where $\delta$ for each system was determined from Fig. 2 and is shown in Table I.

**Figure 4.** Double logarithmic plot of the interatomic distance (defined as distance to half maximum of the intermolecular part of the rdf) vs. density. Both quantities are normalized relative to their respective values at $(\rho_0, T_0)$

Even though isomorph theory predicts that only the molecular center of mass rdf should be invariant. The intermolecular part of the usual atomic radial distribution function (rdf) is also approximately invariant along an isomorph, as shown previously for the LJ system and the three LJ-based molecular liquids[32–38]. In Fig. 3a we examine more carefully the first peak in the intermolecular part of the rdf. The rdf plotted against reduced distance $r\rho^{1/3}$ is an isomorph invariant for the IPL systems (for which this invariance is mathematically exact) but also to an excellent approximation for LJ system. For the molecular systems however, there is a systematic deviation from this structural invariance, with the first peak occurring at shorter reduced $r$ as we move along the isomorph towards higher density and temperature. This may be expected due to the following effect: For an atomic system, moving along an isomorph and decreasing the density, say by a factor $\lambda$, the structure remains invariant and all interatomic distances increase by a factor $\lambda^{1/3}$, the number 3 originating in the number of spatial dimensions. Consider now a molecular system with rigid bonds. For the same volume change, the distances between bonded atoms remain constant, although very generally one would expect the *average* interatomic distance to still be proportional to the volume per particle and scale as $\lambda^{1/3}$. Therefore the distance between *unbonded* atoms must increase by more than $\lambda^{1/3}$ to compensate for the rigid bond structure.

To check this we plot the interatomic distance vs. density along an isomorph in Fig. 4. As a measure of the interatomic distance we choose $r_H$, the distance at which the rdf reaches half of its maximum value. We use this distance and not, for example, the position of the maximum or the most probable distance, to exclude contributions to the rdf from atoms other than nearest neighbors; $r_H$ is also the distance used in ref. 27 to correlate the density scaling exponent with the steepness of the interatomic potential for the atomic LJ system. Plotting $r_H$ as a function of density we see that for the atomic liquid it scales as $\rho^{1/3}$ as expected. However, for the molecular liquids the interatomic distance has a stronger density dependence, which is well described by a power law $\rho^{1/\delta}$, with $\delta < 3$. Returning to Fig. 3(b) and plotting the rdf as a function of $\rho^{1/\delta} r$, we now obtain excellent collapse of the entire first rdf peak.

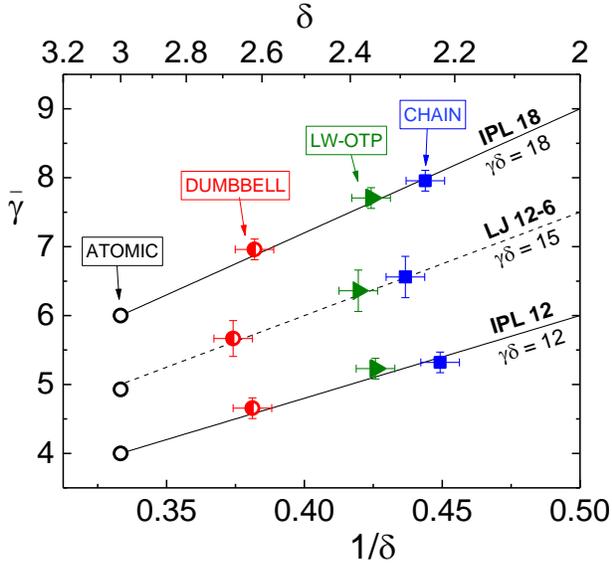 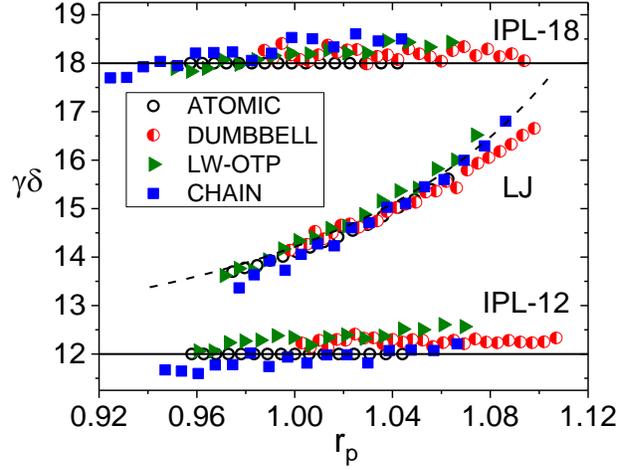

**Figure 5**. Average scaling exponent $\gamma$ plotted against $1/\delta$ for the simulated systems. The lines indicate eq. (8), i.e., the scaling exponent is equal to the steepness of the iteratomic potential divided by the exponent connecting density to intermolecular distance along an isomorph.

**Figure 6.** $\gamma(\rho) \cdot \delta$ plotted against the most probable intermolecular distance for each state point (distance where $r^2 g(r)$ is maximum). The lines indicate the steepness of the interatomic potential. For the LJ potential this is calculated from eq. (9).

From Fig. 4 we make two key observations: First, the exponent $\delta$, describing how intermolecular distance varies with density, depends on the molecular structure and is practically independent of the interatomic potential (LJ, IPL-12 or IPL-18). Second, if we compare to Fig. 1, for a given potential there is an inverse relationship between $\delta$ and the scaling exponent $\gamma$. Plotting the average value of $\gamma$ along the isomorph against $1/\delta$ in Fig. 5, we see not only an excellent correlation, but a quantitative one—the product $\gamma \delta$ is equal to the average steepness of the interatomic potential: 12 for IPL-12 potential, 18 for IPL-18 and $n \sim 15$ for the LJ potential. This suggests the following way to generalize the connection of the intermolecular potential with scaling exponent to molecular systems:

$$\gamma = n/\delta \qquad (8)$$

State point dependence of $\gamma$

For atomic systems, it is possible to accurately estimate $\gamma$ at a given density from the interatomic potential [28,39]. One defines an effective steepness $n(\rho)$ of the potential by fitting an inverse power law in some range of distances near the first rdf peak, with the best results obtained using

$$n(\rho) = -2 - r\frac{u^{(3)}(r)}{u^{(2)}(r)}\bigg|_{r=r_P(\rho)} \qquad (9)$$

where $u^{(n)}(r)$ is the $n$-th derivative of $u(r)$. The distance $r_P(\rho)$ is the most likely intermolecular distance, i.e., the distance where $r^2 g(r)$ is maximum where $g(r)$ is the rdf. Then, the scaling exponent is given by $\gamma(\rho) = n(\rho)/3$ [28].

We can now generalize this to molecular systems: if $\delta$, which depends only on the molecular structure, is known, $n(\rho)$ is calculated in the same way as for atomic systems, $\gamma(\rho) = n(\rho)/\delta$. We test this method in Fig. 6: For each state point along an isomorph we obtain $r_P(\rho)$, and plot $\gamma(\rho)\delta$ as a function of $r_P$. The data for the atomic LJ liquid and three molecular liquids based on the LJ potential collapse to good approximation onto a single curve; moreover, the scaling exponent is very well described by the $n(\rho)$ calculated for the LJ potential.

We have restricted our analysis to molecules based on a single interatomic potential (with variation of the potential parameters, such as the two atom sizes in the asymmetric dumbbell molecule) and rigid bonds. It would be interesting to assess to what extent the results herein can be applied to simulations of more general molecular structures, as well as to experimental results on real liquids. Preliminary simulation results indicate that liquids having harmonic rather than rigid bonds behave similarly, as do angle-dependent potentials, but not those with intramolecular barriers to relaxation (such as dihedral potentials in a polymer)[40].

For an atomic liquid we expect, on general grounds, that the interatomic distance will scale with $\rho^{-1/3}$. The generalization $r_H(\rho) \propto \rho^{-1/\delta}$ works well for the data of Fig. 4, but it is just an empirical fit and may not be general. $\delta$ here plays the role of a dimensionality. For example, we can imagine that when compressing a system of long rigid rod-shaped molecules, while maintaining as much as possible a constant structure, each molecule is compressed in the two lateral dimensions but not the longitudinal one, leading to $\delta \simeq 2$. Indeed, for the molecules studied here $\delta$ seems to be related, loosely speaking, to the aspect ratio of the molecules. It's not clear whether this is the case for different types of interatomic potentials. Consider the S12-6 potential proposed in ref. [41], which unlike the LJ and IPL potentials contains a finite atomic size; the relevant interatomic distance would probably be expected to scale with $(\rho - \rho_0)^{-1/3}$ rather than $\rho^{1/\delta}$.

**SUMMARY**

For molecules based on a single interatomic potential and rigid bonds, we find that each molecular structure (topology, bond lengths, atomic sizes) is associated with an exponent $\delta$ connecting intermolecular distance to density along an isomorph. $\delta$, which compensates for the invariance of molecular bonds to pressure changes in order to fill space and maximize the combinatorial entropy, ranges from 3.0 for atomic liquids to 2.2 for a freely jointed polymer chain. This exponent also links the steepness of the interatomic potential to the density scaling exponent $\gamma$, thus explaining quantitatively why molecules built from the same interatomic potentials exhibit different scaling exponents. This work brings us closer to a theory able to predict the pressure and temperature dependent dynamics of a material from its molecular structure.

## ACKNOWLEDGMENTS

This work was supported by the Office of Naval Research